\documentclass[aps,preprint,tightenlines,floatfix,superscriptaddress,
nofootinbib,showpacs]{revtex4}
\usepackage{graphicx}
\usepackage{dcolumn}
\usepackage{amsfonts}
\usepackage{amsmath}
\def\beq{\begin{equation}}
\def\eeq{\end{equation}}
\def\bea{\begin{eqnarray}}
\def\eea{\end{eqnarray}}
\def\nn{\nonumber}

\def\roughly#1{\mathrel{\raise.3ex\hbox
{$#1$\kern-.75em\lower1ex\hbox{$\sim$}}}}

\def\pslash{p\hspace{-10pt}\not{}\hspace{4pt}}
\def\sslash{s\hspace{-10pt}\not{}\hspace{4pt}}

\def\tbar{{\overline{t}}}
\def\bbar{{\overline{b}}}
\def\dbar{{\overline{d}}}
\def\sbar{{\overline{s}}}
\def\nubar{{\overline{\nu}}}
\def\tbbc{t \to b \bbar c}
\def\ggprocess{gg\to t\tbar\to\left(b\bbar
c\right)\left(\bbar\ell\nubar\right)}
\def\MG{\mbox{MadGraph 5}}

\begin{document}
\bibliographystyle{apsrev}

\preprint{\vbox {\hbox{UdeM-GPP-TH-14-233}}}

\vspace*{2cm}

\title{\boldmath Search for New Physics in Rare Top Decays: \\ $t\tbar$
  Spin Correlations and Other Observables}

\def\tayloru{\affiliation{\it Physics and Engineering Department,
    Taylor University, \\ 236 West Reade Ave., Upland, IN 46989, USA}}
\def\umontreal{\affiliation{\it Physique des Particules, Universit\'e
    de Montr\'eal, \\ C.P. 6128, succ.\ centre-ville, Montr\'eal, QC,
    Canada H3C 3J7}}
\def\laplata{\affiliation{\it IFLP, CONICET -- Dpto. de F\'{\i}sica,
    Universidad Nacional de La Plata, C.C. 67, 1900 La Plata,
    Argentina}}

\tayloru
\umontreal
\laplata

\author{Ken Kiers}
\email{knkiers@taylor.edu}
\tayloru

\author{Pratishruti Saha}
\email{pratishruti.saha@umontreal.ca}
\umontreal

\author{Alejandro Szynkman}
\email{szynkman@fisica.unlp.edu.ar}
\laplata

\author{David London}
\email{london@lps.umontreal.ca}
\umontreal

\author{Samuel Judge}
\email{sdjudge@mtu.edu}
\altaffiliation{Address after 9/1/2014: Department of Mathematics,
Fisher Hall, Michigan Tech University, 1400 Townsend Drive, 
Houghton, Michigan 49931, USA.}
\tayloru

\author{Jordan Melendez}
\email{jordan_melendez@taylor.edu}
\tayloru

\date{\today}

\begin{abstract}
In this paper we study new-physics contributions to the top-quark
decay $t\to b \bbar c$. We search for ways of detecting such new
physics via measurements at the LHC. As top quarks are mainly produced
at the LHC in $t \tbar$ production via gluon fusion, we analyze the
process $\ggprocess$. We find six observables that can be used to
reveal the presence of new physics in $t\to b \bbar c$. Three are
invariant mass-squared distributions involving two of the final-state
particles in the top decay, and three are angular correlations between
the final-state quarks coming from the $t$ decay and the $\ell^-$
coming from the $\tbar$ decay. The angular correlations are related to
the $t \tbar$ spin correlation.
\end{abstract}


\maketitle


\section{Introduction}

Physics beyond the standard model (SM) is expected to exist at
energies above the weak scale. While successive experiments at LEP,
the Tevatron and the LHC have served to validate the SM over the past
few decades, no direct evidence of new physics (NP) has been found
yet. Clearly, NP either exists at an energy scale higher than what has
been probed, or its hints are subtler than we envision. The LHC, which
is currently operational, is essentially a top-quark factory. The
properties of the $t$ can therefore be measured with good precision.
Now, the mass of the top quark is more than an order of magnitude
larger than that of all other fermions. As such, it may be affected by
NP in ways that do not manifest themselves in the interactions of the
lighter fermions. In addition, its large mass causes the top to decay
before it can hadronize, so that it can be studied more or less as a
free quark.

In this paper we study NP contributions to top-quark decay. The
dominant $t$ decay modes in the SM involve $t \to W^+ b$, with $W^+
\to \ell^+ \nu_\ell$, $u \dbar$ or $c \sbar$. Since the
experimentally-measured value of the top decay width is in good
agreement with the SM prediction \cite{SMtopwidth}, it is evident that
the NP contribution to the dominant decay modes, if any, is very small
compared to that of the SM. On the other hand, in the case of decay
modes that are suppressed in the SM, an NP contribution
that is
comparable to that of the SM in that mode may go unnoticed simply
because its impact on the total width is small. This makes it
interesting to probe rare decays, as these could well be where the new
physics is lurking.  One such decay is $t\to W^+ b \to b \bbar c$. It
is suppressed in the SM because it involves the small element $V_{cb}$
($\simeq 0.04$) of the Cabibbo-Kobayashi-Maskawa (CKM) quark mixing
matrix. There are other suppressed decays (e.g., $t\to s \sbar c$),
but here we focus on $t\to b \bbar c$.

Single-top production is rather suppressed at the LHC
\cite{singletprod}, so that it is difficult to isolate the decay
$\tbbc$ experimentally and analyze it on its own. The most significant
production mode for top quarks at the LHC is pair ($t \tbar$)
production. At LHC energies, this is dominated by gluon fusion ($g g
\to t \tbar$), as opposed to quark-antiquark annihilation ($q \bar q
\to t \tbar$). In order to search for NP in top decay, the full
process $gg \to t\tbar$, with $\tbbc$ and $\tbar \to \bbar\ell\nubar$,
must be analyzed. Apart from the usual difficulties of studying a
multi-particle final state, this channel suffers from another
complication -- the $\tbar$ decay leads to a second $\bbar$ in the
final state, providing an additional background that must be taken
into account.

The main purpose of this paper is to analyze the process $gg \to t
(\to b \bbar c) \tbar (\to \bbar\ell\nubar)$, and to look for
observables that can reveal the presence of NP in top
decay\footnote{Here we concentrate on CP-conserving observables. CP
  violation in $t\to b \bbar c$, along the lines of Ref.~\cite{kklrsw},
  will be examined elsewhere \cite{TPselsewhere}.}. We will show that
there are two types of observables that can be used. The first is
simply an invariant mass-squared distribution involving two of the
final-state particles in $\tbbc$. With NP, its form is altered
compared to that of the SM. Note, however, that this type of
observable is entirely related to the decay of the $t$ itself. The
associated production of the $\tbar$ is unimportant, except insofar
that one must distinguish the $\bbar$ quarks coming from the $t$ and
$\tbar$ decays.

The second type of observable does rely on the fact that a $t \tbar$
pair has been produced. The key point is that, in $t \tbar$
production, the spins of the $t$ and $\tbar$ are 
correlated \cite{spincorrexpt}. The spin-correlation coefficient for 
the produced $t \tbar$ pair can be defined as
\begin{equation}
\label{eq:spincorrcoeff}
\kappa_{t \tbar} 
\quad = \quad 
\dfrac{   \sigma_{\uparrow\uparrow} + \sigma_{\downarrow\downarrow}
        - \sigma_{\uparrow\downarrow} - \sigma_{\downarrow\uparrow}
      }
      {   \sigma_{\uparrow\uparrow} + \sigma_{\downarrow\downarrow}
        + \sigma_{\uparrow\downarrow} + \sigma_{\downarrow\uparrow}
      } ~,
\end{equation}
where $\uparrow$ and $\downarrow$ denote the alignment of the spins of
the top and antitop with respect to the chosen spin-quantization axis.
The spin of the $t$ itself is related to the angular distribution of
its decay products through the relation
\begin{equation}
\label{eq:decaydistribution}
\dfrac{1}{\Gamma}\,\dfrac{d\Gamma}{d\cos\chi_i}
\quad = \quad
\dfrac{1}{2}\,(1 + \alpha_i \cos\chi_i) ~,
\end{equation}
where $\chi_i$ is the angle between the direction of the $i^{th}$
decay product and the spin quantization axis in the rest frame of the
top, and $\alpha_i$ is a numerical coefficient whose value depends on
the identity of this decay product.  The spin of the $\tbar$ is
related to the angular distribution of its decay products through a
similar relation, with $\chi_i \to \bar\chi_i$ and $\alpha_i \to
\bar\alpha_i$.  Naturally then, the spin correlation between the
pair-produced top and antitop is manifested in the angular correlation
between the decay products of the two particles. That relation is
given as follows \cite{MP}:
\begin{equation}
\label{eq:angcorr}
\dfrac{1}{\sigma}\,\dfrac{d^2\sigma}{d\cos\chi_i d\cos\bar\chi_j}
\quad = \quad
\dfrac{1}{4}\,
(1 \,+\, \kappa_{t \tbar} \, \alpha_i \, \bar\alpha_j 
                              \cos\chi_i \, \cos\bar\chi_j) ~.
\end{equation}
Its measurement permits the extraction of
$\kappa_{t \tbar}$. If the measured value differs from the prediction
of the SM, it would indicate the presence of NP.

One point should be noted at this juncture. The spin-correlation
coefficient $\kappa_{t \tbar}$ is, by definition, a property of the
$t \tbar$ production process. However, its experimental determination
depends on the decay.  Equation~(\ref{eq:angcorr}) assumes that the $t$ and
$\tbar$ decay via SM interactions only.  If there are NP contributions
in top decay, the value of $\kappa_{t \tbar}$ extracted from the
angular correlations of the top and antitop decay products will be
different from the SM prediction.  This would not be due to a change
in the value of $\kappa_{t \tbar}$ itself, but rather to a change in
the form of Eq.~(\ref{eq:angcorr}).

While there have been several studies of the effect of NP on $t \tbar$
spin correlations, most of them have focused on NP that affects $t
\tbar$ production. These span both CP-conserving~\cite{CPCNP} and
CP-violating~\cite{CPVNP} NP scenarios. Possibilities include
non-standard $g t \tbar$ couplings in the form of anomalous
chromomagnetic dipole or chromoelectric dipole interactions, as well
as many of the NP models proposed to explain the large $t \tbar$
forward-backward asymmetry observed at the Tevatron~\cite{topFBA}.

Of course, NP contributions may be present in both $t \tbar$
production and in the decay. However, NP in the production is much
easier to detect, in that it should be observable even in the dominant
decay modes of the top.  For this reason we ignore the possibility of
NP in $t\tbar$ production in our analysis. We assume it will have
been detected or ruled out before the study of NP in the decay is
done.

Once the observables that carry the signature of NP have been
pinpointed, the next question is: to what extent can they
realistically be used to probe NP in top decay? Can they be used to
identify, even partially, the type of NP present? This is examined in
the companion paper \cite{companion}. There we show that it is likely
that there will be enough events at the LHC to measure these
observables reasonably precisely and extract information about the
nature of NP at play.

In this paper, we begin in Sec.~II by examining how NP in top decay
can affect $\tbbc$.  In Sec.~III we briefly discuss the full pair
production and decay chain $\ggprocess$ (full details are given in the
Appendix). The observables that can be used to search for NP in top
decay are described in Sec.~IV.  In Sec.~V we perform a numerical
simulation of $\ggprocess$ at the LHC, including NP, and compare the
results for the observables with our analytical calculations. We
conclude in Sec.~VI.

\section{New Physics in Top Decay}

As detailed in the introduction, this work focuses on the search for
new physics in rare decays of the top quark.  In this paper, we
examine the decay $\tbbc$. However, the method described here can also
be applied to other suppressed decays such as $t \to s \sbar c$, etc.

While examining a suppressed decay mode, one must consider the most
dominant production mode in order to have sufficient
statistics. Hence, the search for NP in this top decay mode must
involve the process $gg \to t\tbar$. Even there, one may have chosen
to ignore the details of the production process and focus only on the
decay. However, as we show in the following sections, there is
something to be gained by considering the full process $\ggprocess$,
in that the $t\tbar$ spin correlations can be put to use in the
identification of NP.

\subsection{\boldmath $\tbbc$: effective Lagrangian}

In the SM, the decay $\tbbc$ arises via $t\to W^+ b$, followed by $W^+
\to \bbar c$.  NP contributions to $\tbbc$ can be parameterized via an
effective Lagrangian ${\cal L}_{\mbox{\scriptsize eff}}={\cal
  L}_{\mbox{\scriptsize eff}}^V +{\cal L}_{\mbox{\scriptsize
    eff}}^S+{\cal L}_{\mbox{\scriptsize eff}}^T$, with
\begin{eqnarray}
  {\cal L}_{\mbox{\scriptsize eff}}^V & = & 4\sqrt{2}G_F V_{cb}V_{tb}
       \left\{
    X_{LL}^V\,\bbar\gamma_\mu P_L t \,
       \overline{c}\gamma^\mu P_L b
   + X_{LR}^V\,\bbar\gamma_\mu P_L t \,
       \overline{c}\gamma^\mu P_R b
\right.\nonumber\\
& & \hskip2.2truecm \left.
   +~X_{RL}^V\,\bbar\gamma_\mu P_R t \,
       \overline{c}\gamma^\mu P_L b
   + X_{RR}^V\,\bbar\gamma_\mu P_R t \,
       \overline{c}\gamma^\mu P_R b
\right\}+ \mbox{h.c.}, 
\label{eq:eff1}\\
&&\nonumber\\
  {\cal L}_{\mbox{\scriptsize eff}}^S & = & 4\sqrt{2}G_F V_{cb}V_{tb}
\left\{
     X_{LL}^S\,\bbar P_L t \,\overline{c} P_L b
   + X_{LR}^S\,\bbar P_L t \,\overline{c} P_R b
\right. \nonumber\\
& & \hskip2.2truecm \left.
   +~X_{RL}^S\,\bbar P_R t \,\overline{c} P_L b
   + X_{RR}^S\,\bbar P_R t \,\overline{c} P_R b
\right\}+\mbox{h.c.,} 
\label{eq:eff2}\\
&&\nonumber\\
  {\cal L}_{\mbox{\scriptsize eff}}^T & = & 4\sqrt{2}G_F V_{cb}V_{tb}
\left\{
     X^T_{LL} \overline{b}\sigma^{\mu\nu}P_L t \,
     \overline{c}\sigma_{\mu\nu}P_L b 
\right. \nonumber\\
& & \hskip2.2truecm \left.     +~X^{T}_{RR}
\bbar\sigma^{\mu\nu}P_R t \,
     \overline{c}\sigma_{\mu\nu} P_R b
\right\}+\mbox{h.c.}
\label{eq:eff3}
\end{eqnarray}
In the above expressions, colour indices are not shown, but are
assumed to contract in the same manner as those of the SM (i.e., the
fields $\bbar$ with $t$ and $\overline{c}$ with $b$).  In some NP
models, the colour indices would contract in the opposite manner
(i.e., the fields $\overline{c}$ with $t$ and $\bbar$ with $b$).
However, with Fierz transformations it is straightforward to
incorporate colour-mismatched terms into the effective Lagrangian
\cite{kklrsw}.

In general, the NP couplings (the $X$'s in the above
equations) have both weak and strong phases. However, as argued in
Ref.~\cite{DatLon}, since the NP strong phases can only be
generated by self-rescattering from the NP operators, they are very
small. For this reason, we neglect all NP strong phases, so that the
$X$'s contain only weak phases.  Furthermore, the NP couplings can all
reasonably be assumed to be of order unity, so that the SM and NP
contributions to $\tbbc$ can very well be about the same size. When
computing the effect of NP on a particular observable, it is therefore
important to include both the SM-NP and NP-NP interference pieces.

\subsection{\boldmath $\tbbc$: $\left| {\cal M} \right|^2$}

We calculate the square of the matrix element for $\tbbc$ as a
function of the top-quark spin ($s_t$), including the SM and all the
NP contributions. We find 
\begin{eqnarray}
 \frac{1}{3} \!\mathop{
   \sum_{\mbox{\scriptsize colours,}}}_
   {b, \overline{b}, c\mbox{\scriptsize ~spins}}\!\!
   \left| {\cal M}\left(t(s_t)\to b \overline{b}c\right)\right|^2 \!
     & = & 96 G_F^2m_t\left(V_{tb}V_{cb}\right)^2
  \Bigg[
  \sum_{i,\sigma} A_i^\sigma \left(\frac{p_i\cdot p_t}{m_t} - 
  \xi^\sigma p_i\cdot s_t\right) \nonumber \\
   && -16 \,\mbox{Im}\left(X^T_{LL}X^{S*}_{LL}+X^T_{RR}X^{S*}_{RR}\right)
    \epsilon\left(p_t, s_t, p_{\overline{b}},p_c\right) \Bigg] ~,
  \label{eq:ampsquaredtbbbarc}
\end{eqnarray}
where $\epsilon\left(p_t, s_t, p_{\overline{b}},p_c\right) 
\equiv \epsilon_{\mu\nu\rho\sigma} \,
p_t^\mu s_t^\nu p_{\overline{b}}^\rho p_c^\sigma$, 
with $\epsilon_{0123}=-1$ and where $s_t$ is the spin four-vector 
of the top quark. Above, $\sigma = \pm$, $\xi^\pm = \pm 1$ and 
$i=\bbar, b, c$. $A_{\bbar}^+$ is defined as
\begin{eqnarray}
  A_{\bbar}^+ = \left(p_t-p_{\bbar}\right)^2
  \Big[m_W^4\left|G_T\right|^2
  +4 m_W^2\mbox{Re}\left(G_T X^{V*}_{LL}\right)
+\hat{A}_{\bbar}^+\Big] ~,
\label{eq:Abbar}
\end{eqnarray}
where $G_T \equiv G_T(q^2) = (q^2 - M_W^2 + i \Gamma_W M_W)^{-1}$ 
and $q^2 = 2 \,p_{\bbar}\cdot p_c$. The remaining $A_i^\sigma$ are
defined as
\begin{eqnarray}
  A_i^\sigma = \left(p_t-p_{i}\right)^2 \hat{A}_{i}^\sigma ~,
  ~~~~~~~~~~~~~~~~~~~~~~~
  \mbox{(all $i,\sigma$, except $i=\bbar$, $\sigma = +$).}
\end{eqnarray}
In the above,
\begin{eqnarray}
  \hat{A}_{\bbar}^+ &=& 4 \left|X^{V}_{LL}\right|^2
  -8 \,\mbox{Re}\left(X^T_{LL}X^{S*}_{LL}\right)+32 \left|X^T_{LL}\right|^2 ~,
  \nn\\
  \hat{A}_{\bbar}^- &=&
    4\left|X^{V}_{RR}\right|^2
    -8 \,\mbox{Re}\left(X^T_{RR}X^{S*}_{RR}\right)+32 \left|X^T_{RR}\right|^2 ~, \nn\\
  \hat{A}_{b}^+ &=& 
    \left|X^{S}_{LL}\right|^2+\left|X^{S}_{LR}\right|^2
    -16\left|X^{T}_{LL}\right|^2 ~, \nn\\
  \hat{A}_{b}^- &=& 
    \left|X^{S}_{RR}\right|^2+\left|X^{S}_{RL}\right|^2
    -16\left|X^{T}_{RR}\right|^2 ~,\nn\\
  \hat{A}_{c}^+ &=& 
    4\left|X^{V}_{LR}\right|^2
    +8 \,\mbox{Re}\left(X^T_{LL}X^{S*}_{LL}\right)+32\left|X^T_{LL}\right|^2 ~, \nn\\
  \hat{A}_{c}^- &=& 
    4\left|X^{V}_{RL}\right|^2
    +8 \,\mbox{Re}\left(X^T_{RR}X^{S*}_{RR}\right)+32\left|X^T_{RR}\right|^2 ~.
\label{Ahatdefs}
\end{eqnarray}
Note that $A_{\bbar}^+$ contains both the SM and NP contributions,
whereas the other $A_i^\sigma$ contain only NP contributions.

The term proportional to $\epsilon\left(p_t, s_t,
p_{\overline{b}},p_c\right)$ in Eq.~(\ref{eq:ampsquaredtbbbarc})
describes the triple product (TP) in the decay. Because the $X$'s
contain only weak phases, the TP is purely CP-violating. Furthermore,
Eq.~(\ref{eq:Abbar}) contains terms proportional to
$\mbox{Re}(G_T)\mbox{Re}(X^{V*}_{LL})$ and
$\mbox{Im}(G_T)\mbox{Im}(X^{V*}_{LL})$. Of these,
$\mbox{Im}(G_T)\mbox{Im}(X^{V*}_{LL})$ is also CP-violating. Now, if
one adds Eq.~(\ref{eq:ampsquaredtbbbarc}) to its CP-conjugate
counterpart, all CP-violating terms cancel, leaving the remaining
terms unchanged (apart from a normalization factor of $1/2$). In
focusing on CP-conserving observables, we implicitly assume that this
CP averaging has been performed.

The main point to be retained from Eq.~(\ref{eq:ampsquaredtbbbarc}) is
that the amplitude squared depends on seven different combinations of
NP couplings -- six $\hat{A}_{i}^\sigma$'s and Re$(X^{V*}_{LL})$.
Thus, there are a number of independent observables that, in
principle, can provide information about the NP.  While we can hope to
measure all seven of these quantities, we cannot measure all of the
individual $X$ parameters.  In the remainder of this paper (and in the
companion paper), when we refer to ``identifying'' the NP, what is
meant is this partial identification of the six $\hat{A}_{i}^\sigma$'s
and Re$(X^{V*}_{LL})$, not the complete identification of all of the
$X$ parameters.  

\section{\boldmath $\ggprocess$}
\label{Sec:ggprocess}

As a first step, we calculate the cross-section for $t \tbar$ pair
production followed by the decay chain $t \to b \bbar c$, $\tbar \to
\bbar \ell \nu$. We present an outline of the analysis in what
follows; the more technical details can be found in the Appendix.

\begin{figure}[!htbp]
\begin{center}
\resizebox{4in}{!}{\includegraphics*{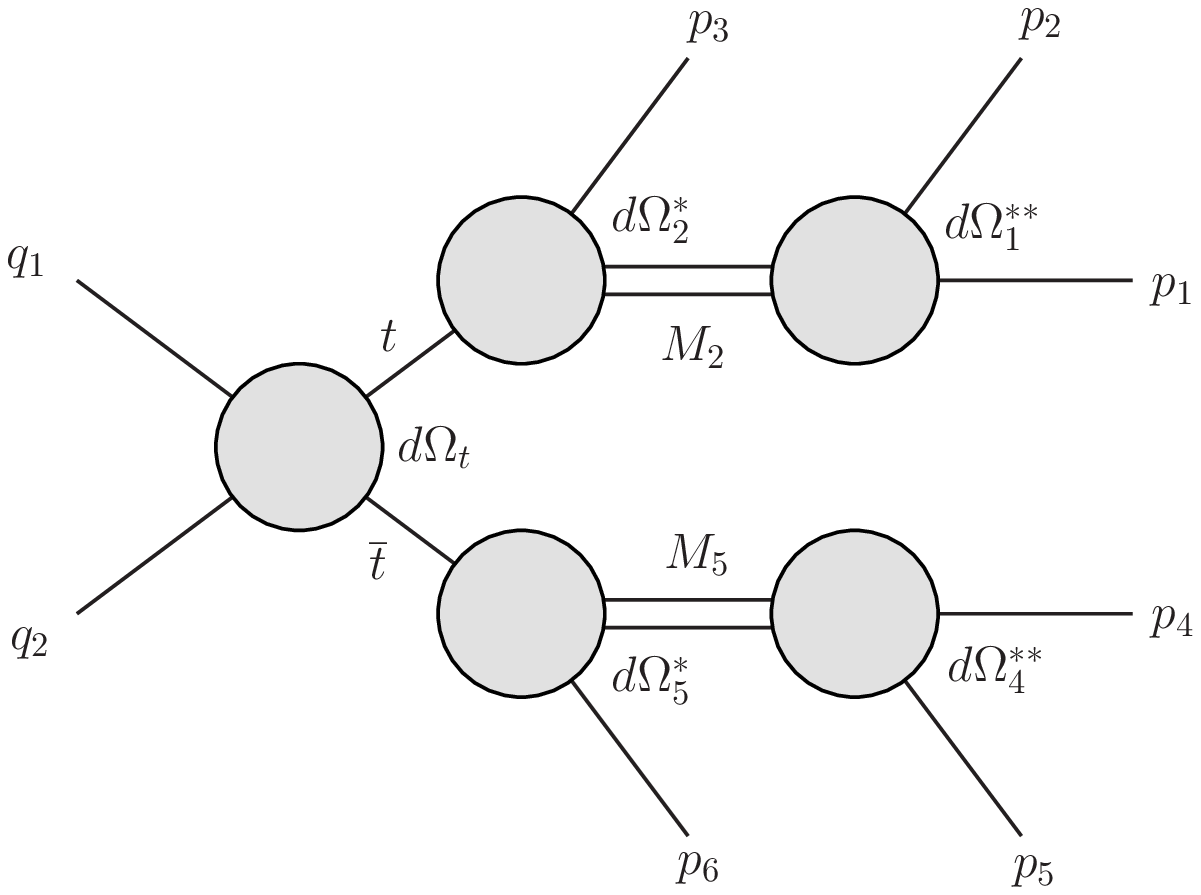}}
\caption{Kinematics for the process $\ggprocess$~\cite{ggprocesskinematics}.
  $\Omega_1^{**}$ denotes the direction of $\vec{p}_1^{~**}$ in the
  rest frame of $M_2$, relative to the direction of
  $\vec{p}_1^{~*}+\vec{p}_2^{~*}$, where $M_2^2 = (p_1+p_2)^2$.
  Similarly, $\Omega_2^*$ denotes the direction of
  $\left(\vec{p}_1^{~*}+\vec{p}_2^{~*}\right)$ in the $t$ rest frame,
  relative to the direction of $\vec{p}_t$ in the $t\tbar$ rest frame.
  $\Omega_t$ denotes the direction of $\vec{p}_t$ relative to
  $\vec{q}_1$, also in the $t\tbar$ rest frame.  The solid angles
  $\Omega_4^{**}$ and $\Omega_5^{*}$ are defined analogously to
  $\Omega_1^{**}$ and $\Omega_2^{*}$, respectively, and $M_5^2 =
  (p_4+p_5)^2$.}
\label{fig:kinematics}
\end{center}
\end{figure}

Briefly, the analysis proceeds as follows.  The process is represented
in Fig.~\ref{fig:kinematics}. The six-body phase space is decomposed
into five solid angles and four invariant masses. The narrow-width
approximation\footnote{The narrow-width approximation is equivalent to
  assuming that the decaying particle is on-shell. Throughout the
  paper, we apply this to the $t$ and $\tbar$ quarks produced via
  gluon fusion, to the $W$ produced in the $\tbar$ decay, and
  generally to the $W$ produced in the $t$ decay.}  is then used for
the $t$ and $\tbar$ quarks to eliminate two of the invariant-mass
degrees of freedom.  The solid angles $d\Omega_1^{**}$,
$d\Omega_2^{*}$, $d\Omega_4^{**}$, $d\Omega_5^{*}$ and $d\Omega_t$ are
defined in five different rest frames, as indicated in
Fig.~\ref{fig:kinematics}. The $*$ and $**$ superscripts indicate that
these angles are defined in reference frames that are, respectively,
one and two boosts away from the $t\tbar$ rest frame. The invariant
masses $M_2$ and $M_5$ are defined through the relations $M_2^2 =
\left(p_1+p_2\right)^2$ and $M_5^2 = \left(p_4+p_5\right)^2$.  In the
end, the differential cross section is a complicated function of the
final-state momenta $p_i$ ($i=1$-6) and the couplings, and is
defined with respect to $dM_2^2 \, dM_5^2 \, d\Omega_1^{**} \,
d\Omega_2^{*} \, d\Omega_4^{**} \, d\Omega_5^{*} \, d\Omega_t$.

We stress that Fig.~\ref{fig:kinematics} represents only the
kinematics of $\ggprocess$. It is not a Feynman diagram. In
particular, $M_5^2$ does not necessarily correspond to the $W^-$
resonance in the $\tbar$ decay, and $M_2^2$ does not necessarily
correspond to the $W^+$ resonance in the SM part of the $t$
decay. Rather, $p_1$, $p_2$ and $p_3$ are the momenta of the $b$,
$\bbar$ and $c$ quarks in $\tbbc$, with all permutations being
allowed. That is, $p_1$, $p_2$ and $p_3$ can each stand for $p_b$,
$p_\bbar$ or $p_c$, and similarly for the particles in the $\tbar$
decay. In constructing the observables, we consider several of these
possibilities.

\section{Observables}
\label{sec:observables}

The first step in finding observables that can yield information about
NP in top decay is to define which final-state particles correspond to
$p_1$-$p_6$. There are several choices possible, corresponding to
different observables.  Throughout this work the momenta for the
$\tbar$ decay products are assigned as follows: $p_4=p_\nubar$,
$p_5=p_\bbar$ and $p_6=p_\ell$.  Taking $p_1 = p_c$, $p_2 = p_b$ and
integrating Eq.~(\ref{eq:dsigma2}) from the Appendix over $M_5^2$ and
over all angles except $\theta_{\bbar}^*$ and $\theta_\ell^*$, we
find\footnote{The angle $\theta_{\bbar}^*$ is ``$\theta_2^*$'' in this
  case (see the caption of Fig.~\ref{fig:kinematics} for a precise
  definition).  This angle is associated with the direction of the
  $b$-$c$ center of mass in the top rest frame, which is opposite to
  the direction of the $\bbar$ in this frame.  Similarly,
  $\theta_\ell^*$ is ``$\theta_5^*$''.}
\begin{eqnarray}
  && \!\!\!\!\!\!\!\!\frac{d\sigma}{d\!\cos\theta_{\bbar}^* 
    \,d\!\cos\theta_\ell^* \,d\zeta_{bc}^2} \nonumber \\
  && =\frac{\sigma_{\mbox{\scriptsize SM}}}{4}
    \Bigg\{\!\!
    \frac{6 \,h_{\mbox{\scriptsize SM}}^{bc}\left(\zeta_{bc}^2\right)}
    {\left(1-\zeta_W^2\right)^2
      \left(1+2\zeta_W^2\right)}
    \left[1 + \kappa(r)
\cos\theta_{\bbar}^*\cos\theta_\ell^*\right]\nonumber\\
  && \!\!\!\!\!\!\!+~\frac{3 G_F m_t^2}
     {\sqrt{2}\pi^2 \left(1-\zeta_W^2\right)^2
      \left(1+2\zeta_W^2\right)}
    \sum_{i,\sigma}
\hat{A}_i^\sigma\left[h_i^{bc}\left(\zeta_{bc}^2\right)
    + \widetilde{h}_i^{bc} \left(\zeta_{bc}^2\right) \xi^\sigma
\kappa(r)
     \cos\theta_{\bbar}^*\cos\theta_\ell^*\right]\!\! \Bigg\},~~
\label{eq:dsigdcosdcosdM}
\end{eqnarray}
where $\zeta_{bc}^2 \equiv \left(p_b+p_c\right)^2/m_t^2$ and
$\zeta_W \equiv m_W/m_t$. $\sigma_{SM}$ is defined in
Eq.~(\ref{eq:sigmaSM}) of the Appendix, the $\hat{A}_i^\sigma$'s are
given in Eq.~(\ref{Ahatdefs}), and $\kappa(r)$ is defined as
\begin{eqnarray}
  \kappa(r) = \frac{\left(-31 r^4 + 37 r^2 -66\right)r 
    -2\left(r^6-17r^4 + 33 r^2-33\right)\tanh^{-1}\left(r\right)}
        {r^2\left[\left(31 r^2 -59\right)r 
          +2\left(r^4-18r^2 + 33\right)\tanh^{-1}\left(r\right)\right]}
~,
\end{eqnarray}
where $r$ is defined in Eq.~(\ref{eq:Pt}) in the Appendix.  Note that
$\langle \kappa(r) \rangle$ = -$\kappa_{t \tbar}$ as defined in
Eq.~(\ref{eq:spincorrcoeff}). The functions $h_i^{bc}(\zeta_{bc}^2)$
and $\widetilde{h}_i^{bc}$ are defined in Table~\ref{tab:hs}, and
\beq
h_{\mbox{\scriptsize SM}}^{bc}\left(\zeta_{bc}^2\right) 
  = \left(1-\zeta_{bc}^2\right)\zeta_{bc}^2\,\,\theta
        (1-\zeta_W^2-\zeta_{bc}^2) ~.
\label{eq:hSM}
\eeq
In writing down Eq.~(\ref{eq:dsigdcosdcosdM}), we have dropped a
contribution proportional to Re$\left(X^{V*}_{LL}\right)$, which tends
to yield a somewhat small effect in practice.  This contribution is
not difficult to calculate, but its inclusion makes the expression for
the differential cross section somewhat cumbersome.  Also, since we
are only considering CP-even contributions, we have dropped a term
proportional to Im$\left(X^{V*}_{LL}\right)$.

\begin{table}[h]
\bgroup
\def\arraystretch{1.2}
\caption{Definitions of the $h_i^{mn}$ ($mn = bc,~\bbar c,~b \bbar$)
  and $\widetilde{h}_i^{bc}$ functions.  The columns correspond to
  $i=b$, $\overline{b}$, $c$.}
\begin{tabular}{c|ccc}
\hline\hline
& $b$ & $\overline{b}$ & $c$ \\
\hline
 $h_i^{bc}(\zeta^2)$ & ~$\frac{1}{2}(1-\zeta^2)^2(1+2\zeta^2)$~ & $3(1-\zeta^2)^2\zeta^2$ 
     & ~$\frac{1}{2}(1-\zeta^2)^2(1+2\zeta^2)$~ \\
$\widetilde{h}_i^{bc}$ & ~$-\frac{1}{2}(1-\zeta^2)^2(1-2\zeta^2)$ & ~$3(1-\zeta^2)^2\zeta^2$ & ~$-\frac{1}{2}(1-\zeta^2)^2(1-2\zeta^2)$ \\
\hline
 $h_i^{\bbar c}(\zeta^2)$ & $3(1-\zeta^2)^2\zeta^2$ & ~$\frac{1}{2}(1-\zeta^2)^2(1+2\zeta^2)$~
     & ~$\frac{1}{2}(1-\zeta^2)^2(1+2\zeta^2)$~ \\ 
\hline
 $h_i^{b \bbar}(\zeta^2)$ & ~$\frac{1}{2}(1-\zeta^2)^2(1+2\zeta^2)$~ & $\frac{1}{2}(1-\zeta^2)^2(1+2\zeta^2)$ 
     & ~$3(1-\zeta^2)^2\zeta^2$ ~ \\ 
\hline\hline
\end{tabular}
\label{tab:hs}
\egroup
\end{table}

Starting from Eq.~(\ref{eq:dsigdcosdcosdM}), we can integrate one or
two more times to obtain differential cross sections in terms of the
two angles or in terms of the invariant mass squared, respectively. 
These are the two types of observables we focus on in this paper: 

{\bf Invariant mass-squared distribution.} Integrating over the
  angles $\theta_{\bbar}^*$ and $\theta_\ell^*$ in
  Eq.~(\ref{eq:dsigdcosdcosdM}) yields
\begin{eqnarray}
  \frac{d\sigma}{d\zeta_{bc}^2} 
  & = & \sigma_{\mbox{\scriptsize SM}}
    \Bigg\{
    \frac{6 \,h_{\mbox{\scriptsize SM}}^{bc}\left(\zeta_{bc}^2\right)}
    {\left(1-\zeta_W^2\right)^2
      \left(1+2\zeta_W^2\right)}\nonumber\\
  && ~~~~~~~~~~~~~+ \frac{3 G_Fm_t^2}
     {\sqrt{2}\pi^2 \left(1-\zeta_W^2\right)^2
      \left(1+2\zeta_W^2\right)}
    \sum_{i,\sigma} \hat{A}_i^\sigma h_i^{bc}\left(\zeta_{bc}^2\right)
\Bigg\}.
\label{eq:dsigdM}
\end{eqnarray}
The above contains three functions of $\zeta_{bc}^2$ that multiply the
various SM and NP terms: $h^{bc}_{\mbox{\scriptsize SM}}$, $h_{\bbar}^{bc}$
and $h_b^{bc}=h_c^{bc}$.  The three functions are qualitatively
different from each other, so that the measurement of the invariant
mass-squared distribution permits the extraction of the NP parameters
$\hat{A}_{\bbar}^+ + \hat{A}_{\bbar}^-$ and $\hat{A}_{b}^+ +
\hat{A}_{b}^- + \hat{A}_{c}^+ + \hat{A}_{c}^-$.

{\bf Angular correlation.} Integrating over $\zeta_{bc}^2$ in
  Eq.~(\ref{eq:dsigdcosdcosdM}), we obtain
\begin{eqnarray}
\label{eq:dsigdcosdcos}
  \frac{d\sigma}{d\!\cos\theta_{\bbar}^* 
    \,d\!\cos\theta_\ell^*}  &=&\frac{\sigma_{\mbox{\scriptsize SM}}}{4}
    \Bigg\{
    \left[1 + \kappa(r)
\cos\theta_{\bbar}^*\cos\theta_\ell^*\right]\nonumber\\
  &+&  \frac{3 G_F m_t^2}
     {4\sqrt{2}\pi^2 \left(1-\zeta_W^2\right)^2
      \left(1+2\zeta_W^2\right)}\Bigg[
    \Bigg(\sum_{i,\sigma} \hat{A}_i^\sigma\Bigg) \\
  &+& \Bigg(\hat{A}_{\bbar}^+-\hat{A}_{\bbar}^-
    -\frac{1}{3}\Big(
      \hat{A}_b^+-\hat{A}_b^-+\hat{A}_c^+-\hat{A}_c^-
    \Big)\Bigg)\kappa(r)
     \cos\theta_{\bbar}^*\cos\theta_\ell^*\Bigg] \Bigg\}. \nn
\end{eqnarray}
By measuring this differential cross section and comparing it to the
SM prediction, one can extract the sum of NP parameters
$\sum_{i,\sigma} \hat{A}_i^\sigma$ and a linear combination of the
differences $\hat{A}_{i}^+ - \hat{A}_{i}^-$ ($i=\bbar, b, c$).  Note
that this observable is sensitive to the $t \tbar$ spin
correlation. For the SM, this is just the coefficient of the term
proportional to $\cos\theta_{\bbar}^*\,\cos\theta_\ell^*$, up to an
overall normalization factor. Once NP is included, this term gets an
additional contribution proportional to a combination of differences
of the NP parameters.

It is straightforward to perform the above analysis for the two other
invariant masses and angles in the $t$ decay.  Taking $p_1 = p_c$
and $p_2 = p_\bbar$, we have
\begin{eqnarray}
  \frac{d\sigma}{d\zeta_{\bbar c}^2} 
  & = & \sigma_{\mbox{\scriptsize SM}}
    \Bigg\{
    \left[1 - 4(1 - \zeta_{\bbar c}^2/\zeta_W^2) \mbox{Re}\!\left(X^{V*}_{LL}\right)\right]
    \frac{6 \,h_{\mbox{\scriptsize SM}}^{\bbar c}\left(\zeta_{\bbar c}^2\right)}
    {\left(1-\zeta_W^2\right)^2
      \left(1+2\zeta_W^2\right)}\nonumber\\
  && ~~~~~~~~~~~~~+ \frac{3 G_Fm_t^2}
     {\sqrt{2}\pi^2 \left(1-\zeta_W^2\right)^2
      \left(1+2\zeta_W^2\right)}
    \sum_{i,\sigma} \hat{A}_i^\sigma h_i^{\bbar c}
       \left(\zeta_{\bbar c}^2\right) \Bigg\},
\label{eq:dsigdM2}
\end{eqnarray}
where the $h_i^{\bbar c}$ are defined in Table~\ref{tab:hs} and
\beq
h_{\mbox{\scriptsize SM}}^{\bbar c}\left(\zeta_{\bbar c}^2\right) 
  = \left(\frac{\zeta_W\gamma_W}{6\pi}\right)
        \frac{(1-\zeta_{\bbar c}^2)^2(1+2\zeta_{\bbar c}^2)}
        {(\zeta_{\bbar c}^2-\zeta_W^2)^2
          +(\zeta_W\gamma_W)^2} ~,
\label{eq:hSMbbarc}
\eeq
with $\gamma_W=\Gamma_W/m_t$.  Here, since $h_b^{\bbar c}$ is
different from $h_\bbar^{\bbar c} = h_c^{\bbar c}$, the measurement of
the invariant mass-squared distribution permits the extraction of the
NP parameters $\hat{A}_b^+ + \hat{A}_b^-$ and $\hat{A}_{\bbar}^+ +
\hat{A}_{\bbar}^- + \hat{A}_{c}^+ + \hat{A}_{c}^-$, as well as
Re$\left(X^{V*}_{LL}\right)$.

The corresponding angular correlation is given by
\begin{eqnarray}
  \frac{d\sigma}{d\!\cos\theta_b^* 
    \,d\!\cos\theta_\ell^*}  &=&\frac{\sigma_{\mbox{\scriptsize SM}}}{4}
    \Bigg\{
    \left[1 + \rho_b(\zeta_W^2)\kappa(r) \cos\theta_b^*\cos\theta_\ell^*\right]\nonumber\\
  &+&  \frac{3 G_F m_t^2}
     {4\sqrt{2}\pi^2 \left(1-\zeta_W^2\right)^2
      \left(1+2\zeta_W^2\right)}\Bigg[
    \Bigg(\sum_{i,\sigma} \hat{A}_i^\sigma\Bigg) \nonumber\\
  && \hskip-2truecm +~\Bigg(\hat{A}_b^+-\hat{A}_b^-
    -\frac{1}{3}\Big(
      \hat{A}_{\bbar}^+-\hat{A}_{\bbar}^-+\hat{A}_c^+-\hat{A}_c^-
    \Big)\Bigg)\kappa(r)
     \cos\theta_b^*\cos\theta_\ell^*\Bigg] \Bigg\} ~,
\label{eq:dsigdcosdcos2}
\end{eqnarray}
where
\begin{eqnarray}
  \rho_b(\zeta_W^2) = -\left(\frac{1-2\zeta_W^2}{1+2\zeta_W^2}\right).
\end{eqnarray}
The measurement of this angular correlation allows one to
extract the sum of NP parameters $\sum_{i,\sigma} \hat{A}_i^\sigma$
and a different linear combination of the differences $\hat{A}_{i}^+ -
\hat{A}_{i}^-$ ($i=\bbar, b, c$) as compared to
Eq.~(\ref{eq:dsigdcosdcos}).

Finally, we take $p_1 = p_b$ and $p_2 = p_{\bbar}$. In this case,

\begin{eqnarray}
  \frac{d\sigma}{d\zeta_{b \bbar}^2} 
  & = & \sigma_{\mbox{\scriptsize SM}}
    \Bigg\{
    \frac{6 \,h_{\mbox{\scriptsize SM}}^{b \bbar}\left(
    \zeta_{b \bbar}^2\right)}
    {\left(1-\zeta_W^2\right)^2
      \left(1+2\zeta_W^2\right)}\nonumber\\
  && ~~~~~~~~~~~~~+ \frac{3 G_Fm_t^2}
     {\sqrt{2}\pi^2 \left(1-\zeta_W^2\right)^2
      \left(1+2\zeta_W^2\right)}
    \sum_{i,\sigma} \hat{A}_i^\sigma h_i^{b \bbar}
       \left(\zeta_{b \bbar}^2\right) \Bigg\} ~,
\label{eq:dsigdM3}
\end{eqnarray}
where the $h_i^{b \bbar}$ are defined in Table~\ref{tab:hs}, and
\beq
  h_{\mbox{\scriptsize SM}}^{b \bbar}
        \left(\zeta_{b\bbar}^2\right) =
        \left(1-\zeta_W^2-\zeta_{b \bbar}^2\right)
         \left(\zeta_W^2+\zeta_{b \bbar}^2\right) 
         \theta\!\left(1-\zeta_W^2-\zeta_{b\bbar}^2\right) ~.
\label{eq:hSMbbarb}
\eeq
We have dropped a contribution proportional to
Re$\left(X^{V*}_{LL}\right)$ in Eq.~(\ref{eq:dsigdM3}), because its
effect is somewhat small in practice.  The measurement of the
invariant mass-squared distribution permits the extraction of the NP
parameters $\hat{A}_c^+ + \hat{A}_c^-$ and $\hat{A}_{b}^+ +
\hat{A}_{b}^- + \hat{A}_{\bbar}^+ + \hat{A}_{\bbar}^-$.

The angular correlation is given by
\begin{eqnarray}
  \frac{d\sigma}{d\!\cos\theta_c^* 
    \,d\!\cos\theta_\ell^*}  &=&\frac{\sigma_{\mbox{\scriptsize SM}}}{4}
    \Bigg\{
    \left[1 + \rho_c(\zeta_W^2)\kappa(r)
\cos\theta_c^*\cos\theta_\ell^*\right]\nonumber\\
  &+&  \frac{3 G_F m_t^2}
     {4\sqrt{2}\pi^2 \left(1-\zeta_W^2\right)^2
      \left(1+2\zeta_W^2\right)}\Bigg[
    \Bigg(\sum_{i,\sigma} \hat{A}_i^\sigma\Bigg) \nonumber\\
  && \hskip-2truecm +~\Bigg(\hat{A}_c^+-\hat{A}_c^-
    -\frac{1}{3}\Big(
   \hat{A}_{b}^+-\hat{A}_{b}^-+\hat{A}_{\bbar}^+-\hat{A}_{\bbar}^-
    \Big)\Bigg)\kappa(r)
     \cos\theta_c^*\cos\theta_\ell^*\Bigg] \Bigg\} \,,
\label{eq:dsigdcosdcos3}
\end{eqnarray}
where
\begin{eqnarray}
\rho_c(\zeta_W^2) = 
\frac{1-12\zeta_W^2+9\zeta_W^4+2\zeta_W^6-12\zeta_W^4\ln(\zeta_W^2)}
     {(1-\zeta_W^2)^2(1+2\zeta_W^2)} \,.
\end{eqnarray}
Here the measurement of the angular correlation allows one to
extract the sum of NP parameters $\sum_{i,\sigma} \hat{A}_i^\sigma$
and a third distinct linear combination of the differences
$\hat{A}_{i}^+ - \hat{A}_{i}^-$ ($i=\bbar, b, c$).

The measurement of any of these observables allows one to detect the
presence of NP in top decay.  If all three angular correlations and
invariant mass-squared distributions can be measured, the results can
be combined to give measurements of all six NP parameters
$\hat{A}_{i}^\sigma$, as well as Re$\left(X^{V*}_{LL}\right)$.
Furthermore, there are numerous measurements, providing significant
redundancy. This is discussed in detail in the companion paper,
Ref.~\cite{companion}.

One can also perform all of the integrations, giving the total cross
section \cite{kklrsw}:
\begin{eqnarray}
  \sigma &=&\sigma_{\mbox{\scriptsize SM}}
    \Bigg\{
 1 + \frac{3 G_F m_t^2}
     {4\sqrt{2}\pi^2 \left(1-\zeta_W^2\right)^2
      \left(1+2\zeta_W^2\right)}
    \sum_{i,\sigma} \hat{A}_i^\sigma\Bigg\}.
\label{eq:sigtot}
\end{eqnarray}
The measurement of $\sigma$ is, in principle, the most straightforward
way to detect NP. Any disagreement between the measured total cross
section and its SM value would indicate NP. The downside of this
approach, however, is that the absolute size of the cross section
might be difficult to determine due to QCD corrections, 
etc.\footnote{Theoretical calculations of the cross section for
  $t{\bar t}$ pair production can be found in Ref.~\cite{topNLO}.
  These include contributions from both $gg$ and $q{\bar q}$, as well
  as higher-order corrections. The $t \tbar$ cross section at the LHC
  at a centre-of-mass energy of 14 TeV is $\sim 900$ pb.} For this
reason it may be better to use the invariant mass-squared
distributions and/or the angular correlations.

The measurement of the triple-differential distribution of
Eq.~(\ref{eq:dsigdcosdcosdM}) would give a great deal of information
about the NP parameters. However, it is unlikely there will be
sufficient statistics to allow this measurement to be carried out.

\section{Numerical Simulation}

The expressions in the previous section provide a clear picture of the
corrections to the various observables introduced by the new-physics
contributions. In order to obtain meaningful projections in the
context of the LHC, we perform a numerical simulation using
\MG\ \cite{MG5}.  The new couplings due to the effective Lagrangian
[Eqs.~(\ref{eq:eff1})-(\ref{eq:eff3})] are incorporated into
\MG\ via FeynRules \cite{FeynRules}. We compute $gg \to t
\tbar \to (b \bbar c) \, (\bbar e^- \nubar_e)$ and obtain
the $d\sigma/d\zeta^2_{ij}$ distributions and the angular
correlations discussed in the previous section. 
This naturally involves the convolution of the cross sections and
differential cross sections calculated at the parton level with the
appropriate parton densities. We use CTEQ6L1 PDFs \cite{PDFs} with 
the factorization and renormalization scales set to $m_t$ = 172 GeV.

\begin{figure}[!htbp]
\begin{center}
\resizebox{5in}{!}{\includegraphics*{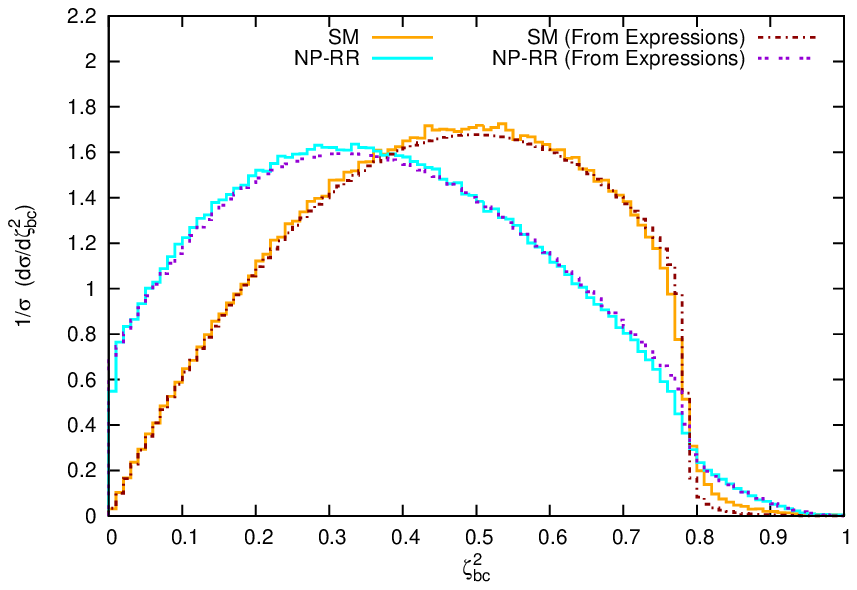}}
\caption{A comparison of the normalized $d\sigma/d\zeta^2_{bc}$
  distribution obtained using \MG\ with that of
  Eq.~(\ref{eq:dsigdM}). This is done for the SM and for the NP
  scenario in which $X^S_{RR} = X^V_{RR} = X^T_{RR} = 1 + i$, with all
  other NP parameters set to zero (labelled here as NP-RR).}
\label{fig:zetasq_bc}
\end{center}
\end{figure}

In Fig.~\ref{fig:zetasq_bc}, we compare the normalized
$d\sigma/d\zeta^2_{bc}$ distribution obtained using \MG\ for a $pp$
collider with a centre-of-mass energy of 14 TeV with that obtained for
gluons colliding at a fixed centre-of-mass energy of 600 GeV using the
analytic expression in Eq.~(\ref{eq:dsigdM}). This is done for both
the SM and a particular NP scenario\footnote{We take $X^S_{RR} =
  X^V_{RR} = X^T_{RR} = 1 + i$.  Note that, although these NP
  parameters are complex, there are no SM-NP or NP-NP interference
  effects. As such, they do not lead to CP violation.}. In both cases
there is remarkable agreement between the two methods of obtaining
$d\sigma/d\zeta^2_{bc}$.  At first glance, this may seem extremely
surprising, but a slightly closer look at the issue reveals that it is
not really so.

The $d\sigma/d\zeta^2_{ij}$ distributions involve only the decay
products of the top. Any observable that involves only particles
coming from a single decay can be computed in the rest frame of the
decaying particle and converted to its laboratory-frame equivalent by
applying a Lorentz boost. At a $pp$ collider, each event would be
associated with a different boost. But since $\zeta^2_{ij}$ is
Lorentz-invariant by construction, the distributions can be expected
to look identical in both the top rest frame and the laboratory frame,
which is what is seen in Fig.~\ref{fig:zetasq_bc}.

Note, however, that the observed $d\sigma/d\zeta^2_{ij}$ distribution
is the result of an ensemble of top decays in which the top quarks are
not all identical to begin with. While most of the top quarks are
produced on-shell, the ensemble also includes top quarks that are
off-shell to varying degrees.  Moreover, the virtuality of the tops is
distributed differently in the fixed-energy and variable-energy cases:
in the fixed-energy case one has the additional condition that $(p_t +
p_{\tbar})^2$ is fixed. Nevertheless, it turns out that this is a
small effect. The normalized distributions for the two cases look
almost identical, and the inclusion of PDFs does not lead to any
significant change in their shape. The slight (noticeable) difference
in the region $\zeta^2_{bc} \approx$ 0.8 is due to the following. In the
analytic expressions, the widths of the $t$ and the $W$ are dealt with
in slightly different ways.  For the $t$, the narrow-width
approximation is incorporated by making the substitution (see the Appendix)
\begin{equation}
\dfrac{1}{(p_t^2 - m_t^2)^2 + \Gamma_t^2\,m_t^2} 
\quad \longrightarrow \quad
\dfrac{\pi}{\Gamma_t\,m_t}\,\delta(p_t^2 - m_t^2) \, .
\end{equation}
For the $W$, the result of applying the narrow-width approximation is
encapsulated in the factor $\theta(1 - \zeta_W^2 - \zeta^2_{bc})$
appearing in the definition of $h^{bc}_{\mbox{\scriptsize SM}}$ in
Eq.~(\ref{eq:hSM}). The finite width of the $W$ can be approximated to
some extent by making the replacement
\begin{equation}
\theta(1 - \zeta_W^2 -\zeta^2_{bc}) 
\quad \longrightarrow \quad
\dfrac{1}{\pi} \, 
\left[ 
\tan^{-1}
\left(\dfrac{1 - \zeta_W^2 -\zeta^2_{bc}}{\zeta_W\gamma_W}\right)
\,+\,
\tan^{-1}
\left(\dfrac{\zeta_W}{\gamma_W}\right)
\right] \, .
\end{equation}
This is included in the curves in Fig.~\ref{fig:zetasq_bc}.  On the
other hand, in \MG\ both the $t$ and $W$ widths are dealt with
identically with the integral covering an interval of $m \pm 15\Gamma$
in each case\footnote{Within \MG, this is governed by the parameter
  \texttt{bwcutoff}, which takes the default value 15.}.

\begin{figure}[!htbp]
\begin{center}
\resizebox{6in}{!}{\includegraphics*{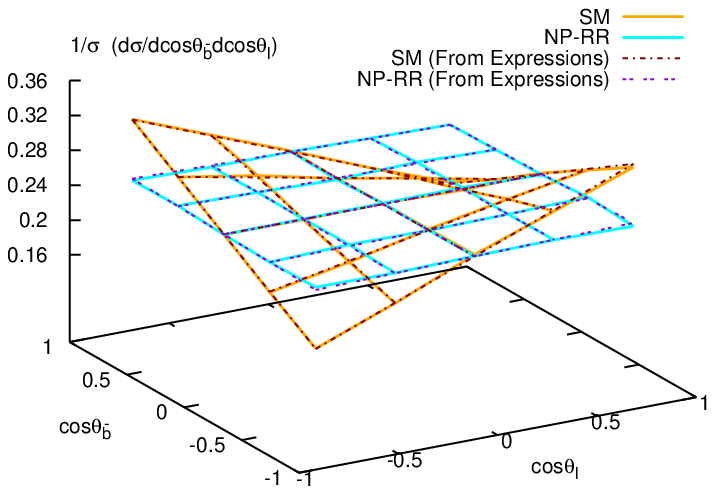}}
\caption{A comparison of the normalized $d\sigma/d\cos\theta_\bbar^*
  d\cos\theta_\ell^*$ angular distribution obtained using \MG\ with that
  of Eq.~(\ref{eq:dsigdcosdcos}). This is done for the SM and for the
  NP scenario in which $X^S_{RR} = X^V_{RR} = X^T_{RR} = 1 + i$
  (labelled here as NP-RR).}
\label{fig:ctheta_bbarl}
\end{center}
\end{figure}

Next we turn to the angular correlation (Fig.~\ref{fig:ctheta_bbarl}).
This observable involves decay products coming from two different
decays. It is therefore sensitive not only to the physics in the two
decays (whether new or standard), but also to the correlations in the
production of the two decaying particles (i.e., the $t \tbar$ spin
correlations). This information is contained in the factor
$\kappa(r)$.  In a fixed-energy gluon-gluon collison, $\kappa(r)$ is
fixed.  In our expression [Eq.~(\ref{eq:dsigdcosdcos})], $\kappa(r)$
is replaced by its expectation value $\langle \kappa(r) \rangle$. When
this is calculated over the energy range sampled in 14 TeV $pp$
collisions, we find that, once again, the normalized distributions
obtained using this expression agree very well with those obtained
from the full numerical simulation using \MG.

The fact that the analytical expressions for the observables agree
with numerical simulations suggests that it is possible to extract
some of the new-physics parameters by fitting the shapes of these
distributions. We present the results of these fits in the companion
paper \cite{companion}. Note that, in comparing the analytical
expressions with the \MG\ simulation, we have taken the $\bbar$ quark
to be that coming from the $t$ decay. However, as noted in the
introduction, there is also a $\bbar$ coming from the $\tbar$ decay,
and this background must be taken into account. This issue, along with
other complications, is addressed in Ref.~\cite{companion}.

\section{Conclusions}

In this paper we study new-physics (NP) contributions to top-quark
decay. Such effects can be significant only for decays that are
suppressed in the SM. Here we focus on $t\to b \bbar c$, whose SM
amplitude involves the small element $V_{cb}$ ($\simeq 0.04$) of the
CKM matrix. Allowing for all Lorentz structures, there are ten
possible dimension-6 NP operators that can contribute to this decay. 
The goal is to find ways of detecting the presence of such NP in
$\tbbc$.

Since the LHC produces top quarks copiously, it is an excellent place
to search for signals of NP in $t\to b \bbar c$. However, the dominant
mode for top-quark production is pair ($t \tbar$) production via gluon
fusion: $g g \to t \tbar$. This makes it difficult to study $\tbbc$ on
its own. In order to search for NP in top decay, the full process
$\ggprocess$ must be analyzed.

We consider only CP-conserving NP, and find that there are two types
of observables that can be used to reveal the presence of NP in top
decay. The first is an invariant mass-squared distribution involving
two of the final-state particles in $\tbbc$. There are three such
distributions.  The second is an angular correlation between the decay
products of the $t$ and $\tbar$. This is related to the $t \tbar$ spin
correlation. We consider the angular correlation between one of the
final-state quarks in $\tbbc$ and the $\ell^-$ coming from the $\tbar$
decay. There are three such correlations. The six observables depend
on different combinations of the coefficients of the ten NP operators.

We compare the analytical expressions for the observables with the
results of a numerical simulation of the LHC using \MG. We find that
the agreement between the two is excellent. This suggests that the
measurement of these observables can indeed be used to extract some of
the new-physics parameters. In the companion paper,
Ref.~\cite{companion}, we demonstrate this explicitly by performing
fits of such measurements. We also show how to deal with complications
such as the background due to the $\bbar$ coming from the $\tbar$
decay.

\bigskip
\noindent
{\bf Acknowledgments}: The authors wish to thank the MadGraph and
FeynRules Teams for extensive discussions about MadGraph and
FeynRules, respectively.  The authors are also indebted to German
Valencia and Howard Baer for helpful discussions and to Zach Bethel
and Carl Daudt for technical support.  This work was financially
supported by NSERC of Canada (DL, PS).  In addition, this work has
been partially supported by ANPCyT under grant No. PICT-PRH 2009-0054
and by CONICET (AS).  The work of SJ and JM was supported by the
U.S.\ National Science Foundation under Grant PHY--1215785.  The work
of KK was supported by the U.S.\ National Science Foundation under
Grants PHY--0900914 and PHY--1215785.  KK also acknowledges sabbatical
support from Taylor University.

\section*{Appendix}
\appendix

In this Appendix we work out an expression for the differential cross
section for $\ggprocess$.  Our main result may be found below in
Eqs.~(\ref{eq:dsigma2})-(\ref{eq:dlambda}).  As an intermediate step,
we write the differential cross section for $\ggprocess$ in a
quasi-factorized form that makes use of expressions for $gg\to t
\tbar$, $t\to b\bbar c$ and $\tbar\to \bbar\ell\nubar$ (see
Eq.~(\ref{eq:diffcrosssectiondoublespin2}), below).  Throughout, we
assume that NP is present only in $\tbbc$; $gg\to t\tbar$ and
$\tbar\to \bbar\ell\nubar$ are purely SM in nature.  Furthermore, we
always employ the narrow-width approximation for the $t$ and $\tbar$,
which is equivalent to assuming that they are on-shell.

\subsection{\boldmath $g g \to t \tbar$}
\label{sec:ggttbar}

\begin{figure}[!htbp]
\begin{center}
\resizebox{5in}{!}{\includegraphics*{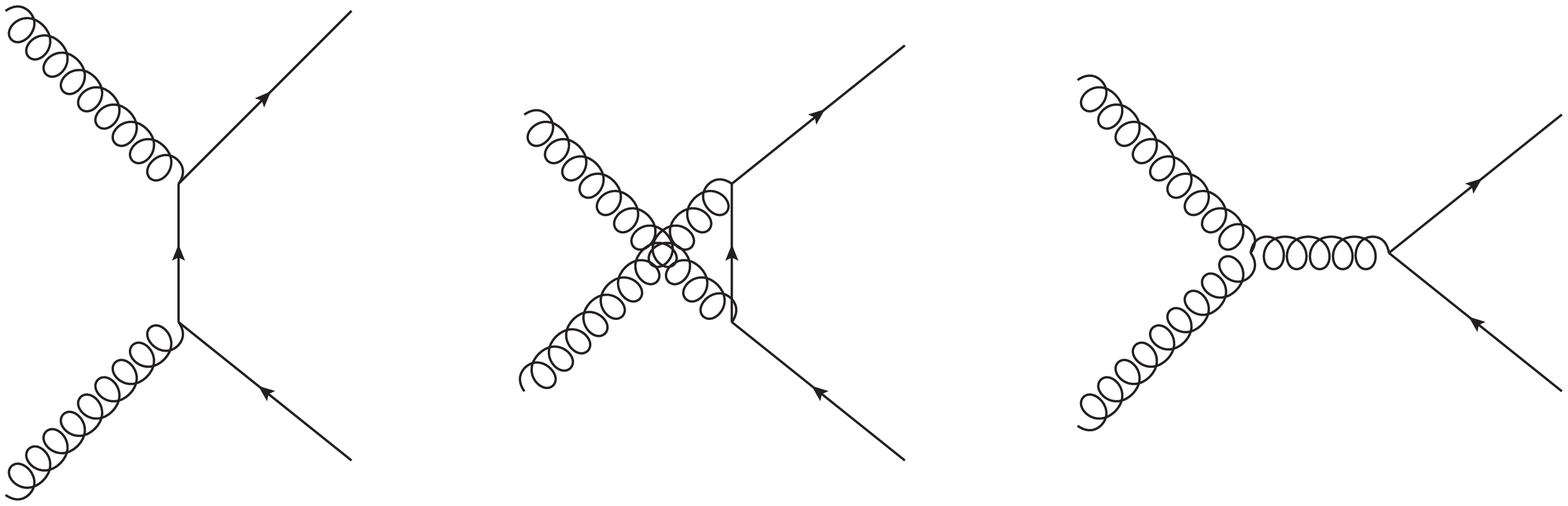}}
\caption{Feynman diagrams for $gg\to t \tbar$.  The $t$
subsequently decays to $b\bbar c$.}
\label{fig:gluonfusion}
\end{center}
\end{figure}

We begin with $g g \to t \tbar$ (see Fig.~\ref{fig:gluonfusion}). The
amplitude squared for $gg\to t\tbar$, including the $t$ and $\tbar$
polarizations, was computed in Ref.~\cite{soni_ggttbar}.  It is useful
to define the following quantities:\footnote{These definitions are
  slightly different from those used in Ref.~\cite{soni_ggttbar}.}
\begin{eqnarray}
  & P_t \equiv p_t-p_{\tbar} ~,~~
  Q \equiv q_1+q_2 = p_t+p_{\tbar} ~,~~
  P_g \equiv q_1-q_2 ~, & \nn\\
  & r \equiv \sqrt{1 - 4 m_t^2/Q^2} ~,~~  
  z \equiv - P_t \cdot P_g/(r Q^2) ~, &
\label{eq:Pt}
\end{eqnarray}
where $p_t$ and $p_{\tbar}$ are the $t$ and $\tbar$ momenta, and $q_1$
and $q_2$ are the momenta of the initial gluons.  The matrix element
squared is then given by (see also Fig.~\ref{fig:gg->4and6fermions}
and Eq.~(\ref{eq:Mabij}) below),
\begin{eqnarray}
  &&\!\!\!\!\!\!\!\!\!\!\!\!\!
   \frac{1}{256}\mathop{\sum_{a,b,i,j;}}_
       {\mbox{\scriptsize gluon pol'ns}} 
    \left|{\cal M}^{ab,ij}\left(gg\to t(s_t)\tbar\left(s_\tbar\right)\right)\right|^2 \nonumber \\
      & = & \frac{g_s^4 (9r^2z^2+7)}{192 (r^2z^2-1)^2}
    \Big\{
    - f(r,z) + s_t\cdot s_{\tbar} \, g(r,z) \nonumber\\
    && \hskip-0.5truecm
+~\frac{r^2(r^2-1)(z^2-1)}{2 m_t^2}\left[
      P_g\!\cdot\! s_t \left(
       P_g\!\cdot\! s_{\tbar} - Q\!\cdot\! s_{\tbar}\,rz
      \right) +
      Q\!\cdot\! s_t \left(
       P_g\!\cdot\! s_{\tbar}\,rz - Q\!\cdot\! s_{\tbar}
      \right)
      \right]
    \Big\} ~,
    \label{eq:ggttbar}
\end{eqnarray}
in which
\begin{eqnarray}
  f(r,z) & = & z^4r^4+2r^2z^2\left(1-r^2\right)+2r^4-2r^2-1 ~,
  \label{eq:f} \\
  g(r,z) & = & r^4\left(z^4 -2z^2+2\right)-2r^2+1 ~.
  \label{eq:g}
\end{eqnarray}
Integrating the amplitude squared over phase space and summing over the
$t$ and $\tbar$ spins yields the following expression for the
parton-level scattering cross-section:
\begin{eqnarray}
\sigma \left(gg\to t\tbar\right) 
     \!& = & \!\frac{\pi \alpha_s^2 (1-r^2)}{192 m_t^2}
     \left[r(31 r^2-59) + 2 (r^4-18 r^2+33)\tanh^{-1}(r)\right] ~.
\label{eq:sigggttbar}
\end{eqnarray}

\subsection{\boldmath Formal Factorization of the Production and
Decay Processes.}
\label{sec:formal}
\begin{figure}[t]
\begin{center}
\resizebox{5.9in}{!}{\includegraphics*{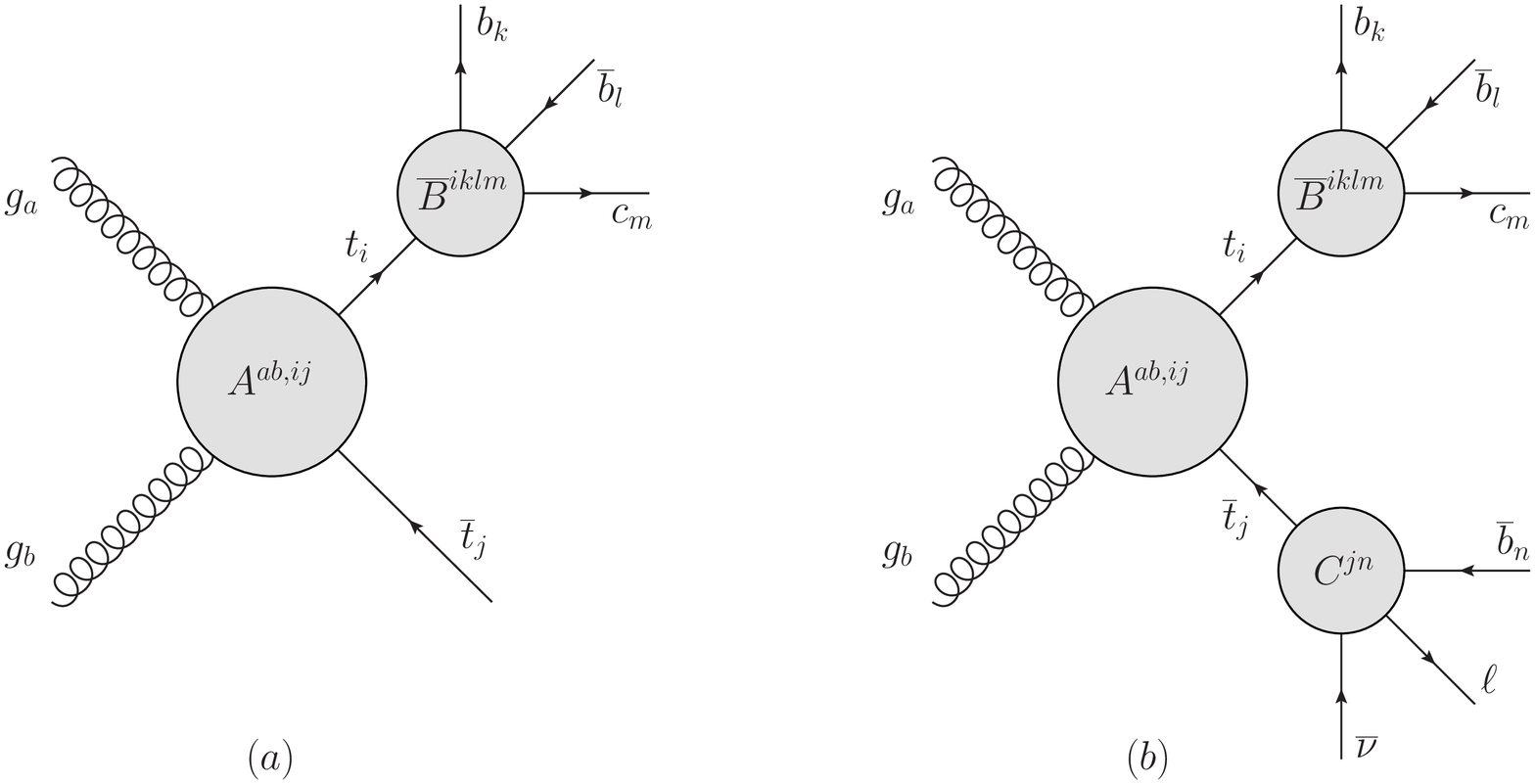}}
\caption{Feynman diagrams for $gg\to t \tbar\to
  \left(b\bbar c\right)\tbar$ and $gg\to t
  \tbar\to
  \left(b\bbar c\right)\left(\bbar\ell\nubar\right)$.
  The subscripts $i, j,\ldots, n$ are colour indices.  All NP effects
  are assumed to be contained in ${\overline{B}}$; $gg\to t\tbar$ and
$\tbar\to \bbar\ell\nubar$ are assumed to be SM-like.}
\label{fig:gg->4and6fermions}
\end{center}
\end{figure}

We now derive expressions that can be used to translate
$t$-spin-dependent observables into a form that may be more useful to
experimentalists.  Our starting point is the observation that the spins
of the $t$ and the $\tbar$ are correlated in $gg\to t \tbar$ [see
  Eq.~(\ref{eq:ggttbar})].  Thus, $t$-spin observables can in
principle be translated into observables that employ the spin of the
$\tbar$.  This is shown schematically in
Fig.~\ref{fig:gg->4and6fermions}~(a).  Of course, the spin of the
$\tbar$ is itself not directly measurable.  Fortunately, however, the
momentum of the charged lepton in $\tbar\to \bbar \ell \nubar$ is
correlated with the spin of the $\tbar$.  Thus, in order to consider
$t$-spin-dependent observables in $t\to b \bbar c$, we can study the
full process $\ggprocess$, as is indicated in
Fig.~\ref{fig:gg->4and6fermions}~(b).

Consider the two diagrams shown in Fig.~\ref{fig:gg->4and6fermions}.
The matrices $A$, ${\overline{B}}$ and $C$ indicated there are defined
via the production and decay amplitudes as follows:
\begin{eqnarray}
  {\cal M}^{ab,ij}
    \left(gg\to t(s_t)\tbar\left(s_{\tbar}\right)\right) 
    & = & \overline{u}_t\left(p_t, s_t\right) A^{ab,ij}
      v_{\tbar}\left(p_{\tbar},s_{\tbar}\right)\, ,
  \label{eq:Mabij}\\
  {\cal M}^{iklm}
    \left(t(s_t)\to b \bbar c\right)
    & = & {\overline{B}}^{iklm} u_t\left(p_t, s_t\right) \,, \\
  {\cal M}^{jn}
    \left(\tbar\left(s_{\tbar}\right)\to
    \bbar\ell\nubar\right)
    & = & \overline{v}_{\tbar}
      \left(p_{\tbar},s_{\tbar}\right)C^{jn} \,,
\end{eqnarray}
in which $i,j,\ldots, n$ are colour indices.  We assume that colour
indices contract as in the SM, so that
\begin{eqnarray}
  {\overline{B}}^{iklm}  = {\overline{B}}\, \delta_{ik}\delta_{lm}
~~~\mbox{and}~~~
  C^{jn} = C\, \delta_{jn} \, .
\end{eqnarray}
Note that $\overline{B}$ is assumed to contain all of the NP effects.
Explicit calculation, starting from the effective Lagrangian given in
Eqs~(\ref{eq:eff1})-(\ref{eq:eff3}), yields
\begin{eqnarray}
  {\overline{B}} & = & 4\sqrt{2}G_FV_{cb}V_{tb}\Bigg[\frac{1}{2}m_W^2 
    \left(\overline{u}_b\gamma_\mu P_L\right) \left(\overline{u}_c\gamma^\mu P_L v_{\bbar}\right)
    G_T\left(2p_{\bbar}\cdot p_c\right) \nonumber\\
    &&  ~~~~~~~~~~~~~~~~+
    X^V_{LL} \left(\overline{u}_b\gamma_\mu P_L\right) \left(\overline{u}_c\gamma^\mu P_L v_{\bbar}\right)
    + X^V_{LR} \left(\overline{u}_b\gamma_\mu P_L\right) \left(\overline{u}_c\gamma^\mu P_R v_{\bbar}\right)
    +\ldots. \Bigg]
\end{eqnarray}
Furthermore, we define $\overline{A}$, $B$ and $\overline{C}$ via the
following relations
\begin{eqnarray}
  \overline{A}^{ab,ij} \equiv \gamma^0
\left(A^{ab,ij}\right)^\dagger\gamma^0,
  ~~~{\overline{B}} \equiv B^\dagger \gamma^0
  ~~~\mbox{and}~~~\overline{C} \equiv C^\dagger \gamma^0 \, .
\end{eqnarray}

Let us begin by considering the diagram in
Fig.~\ref{fig:gg->4and6fermions}~(a).
The amplitude for this process may be written as follows
\begin{eqnarray}
  {\cal M}^{ab,klmj}\!\!\left(gg\to \!\left(b\bbar c\right)\!
     \tbar\left(s_{\tbar}\right)\!\right)\!\!
  = -\frac{1}{p_t^2-m_t^2+i\Gamma_t m_t}\sum_i{\overline{B}}^{iklm}\!\!
  \left(\pslash_t+m_t\right)\!A^{ab,ij}v_{\tbar}\!
    \left(p_{\tbar},s_{\tbar}\right) ~.
\end{eqnarray}
Multiplying the above expression by its complex conjugate and making the
substitution
\begin{eqnarray}
  \left[\left(p_t^2-m_t^2\right)^2+ \Gamma_t^2 m_t^2\right]^{-1}
  \simeq \frac{\pi}{\Gamma_t m_t}\delta\!\!\left(p_t^2-m_t^2\right) ~,
\end{eqnarray}
we have
\begin{eqnarray}
  &&
\!\!\!\!\!\!\!\!\!\!\frac{1}{256}\sum_{a,b,k,l,m,j}\,\sum_{\mbox{
\scriptsize spins}}
   \left|{\cal M}^{ab,klmj}
  \left(gg\to \left(b\bbar c\right)
     \tbar\left(s_{\tbar}\right)\right)\right|^2 \\
  && =\frac{3\pi}{256\,\Gamma_t m_t}
     \sum_{a,b,k,j}\,\sum_{\mbox{\scriptsize spins}}
     \delta\!\!\left(p_t^2-m_t^2\right) \nonumber \\
  && ~~\times\frac{1}{2}
     \mbox{Tr}\left[B{\overline{B}}\left(\pslash_t+m_t\right)A^{ab,kj}
       \left(\pslash_{\tbar}-m_t\right)
       \left(1+\gamma^5\sslash_{\tbar}\right)
       \overline{A}^{ab,kj}\left(\pslash_t+m_t\right)\right]\,,
\label{eq:LargeTraceSingleSpin}
\end{eqnarray}
where the sum over spins includes the gluon spins as well as those of
the $b$, $\bbar$ and $c$.  Note that the $t$ quark only shows
up via its propagator in this expression.  Thus, the spin of the $t$
is summed over, as it should be.  The spin of the $\tbar$, however,
appears explicitly.

This can be simplified by using the following identity, which is
similar to an expression in Ref.~\cite{kawasaki} (see also
Ref.~\cite{arens}):
\begin{eqnarray}
  \mbox{Tr}\left[\mathbb{X}\left(\pslash \pm
m\right)\mathbb{Y}\left(\pslash \pm m \right)\right]
   & = & \frac{1}{2}\Big\{\mbox{Tr}\left[\mathbb{X}\left(\pslash \pm
m\right)\right]
      \mbox{Tr}\left[\mathbb{Y}\left(\pslash \pm m\right)\right]
\nonumber \\
   && \hskip-1truecm
-~\eta_{\mu\nu}\mbox{Tr}\left[\mathbb{X}\left(\pslash \pm
m\right)\gamma^5\gamma^\mu\right]
      \mbox{Tr}\left[\mathbb{Y}\left(\pslash \pm
m\right)\gamma^5\gamma^\nu\right]
    \Big\} ~,
\label{eq:XYtrick}
\end{eqnarray}
where
\begin{eqnarray}
  \eta_{\mu\nu} \equiv g_{\mu\nu} - \frac{p_\mu p_\nu}{m^2} ~,
\end{eqnarray}
and where it is assumed that $p^2=m^2$.  Setting
\begin{eqnarray}
  \mathbb{X} & = & B{\overline{B}} ~, \\
  \mathbb{Y} & = & \frac{1}{2}A^{ab,kj}
       \left(\pslash_{\tbar}-m_t\right)
       \left(1+\gamma^5\sslash_{\tbar}\right)
       \overline{A}^{ab,kj} ~,
\end{eqnarray}
we can split the trace in Eq.~(\ref{eq:LargeTraceSingleSpin}) into two
pieces, one corresponding to the $t\tbar$ production ($\mathbb{Y}$)
and one to the $t$ decay ($\mathbb{X}$).  Finally, defining
\begin{eqnarray}
  n_{t\mu} \equiv -\eta_{\mu\nu}\mbox{Tr}\!\left[B{\overline{B}}
    \left(\pslash_t + m_t\right)\gamma^5\gamma^\nu\right]/
     \mbox{Tr}\!\left[B{\overline{B}}\left(\pslash_t +
m_t\right)\right],
\label{eq:nt}
\end{eqnarray}
we find that we can write the differential cross section corresponding
to Fig.~\ref{fig:gg->4and6fermions}~(a) in the following suggestive
form~\cite{kawasaki}:
\begin{eqnarray}
  d\sigma\left(gg\to \left(b\bbar c\right)
    \tbar\left(s_{\tbar}\right)\right)
   = \frac{2}{\Gamma_t}\sum_{b, \bbar, c \mbox{\scriptsize ~spins}}
     d\sigma\left(gg\to t\left(n_t\right)
       \tbar\left(s_{\tbar}\right)\right)
     d\Gamma\left(t\to b \bbar c\right) \, .
\label{eq:diffcrosssectionsinglespin}
\end{eqnarray}

Note, however, that there are a few subtleties involved in writing the
differential cross section in this way.  In particular, 
\begin{enumerate}
\item The $t$ polarization, $n_{t\mu}$, is a very particular
  four-vector, defined in Eq.~(\ref{eq:nt}).
\item While $d\sigma\left(gg\to t\left(n_t\right)
  \tbar\left(s_{\tbar}\right)\right)$ is calculated for
  a particular spin four-vector for the $t$, the $t$ spin is averaged
  in $d\Gamma\left(t\to b \bbar c\right)$.
\item Although Eq.~(\ref{eq:diffcrosssectionsinglespin}) has the
  appearance of being factorized cleanly into two pieces, the
  $t$-polarization four-vector contained in $d\sigma\left(gg\to
  t\left(n_t\right) \tbar\left(s_{\tbar}\right)\right)$ depends on the
  phase-space variables contained in $d\Gamma\left(t\to b \bbar
  c\right)$.  Similarly, the spin four-vectors for the $b$, $\bbar$
  and $c$ appear both in $n_{t\mu}$ and in $d\Gamma\left(t\to b \bbar
  c\right)$.
\item Given the preceding comment, one must exercise some caution when
  integrating over phase space and summing over the $b$,
  $\bbar$ and $c$ spins.  In particular, one must do so for the
  product of $d\sigma(gg\to t\tbar)$ and $d\Gamma\left(t\to b
  \bbar c\right)$, and not for the two quantities separately.
  For the spin sum, the $t$-polarization-dependent quantity that
  appears in calculations is always $\mbox{Tr}\!\left[B\overline{B}\left(\pslash_t
    + m_t\right)\right] n_{t\mu}$.  It is safe to sum this quantity
  over spins.
\end{enumerate}

The above approach can be generalized to the scenario indicated in
Fig.~\ref{fig:gg->4and6fermions}~(b) by applying the trick in
Eq.~(\ref{eq:XYtrick}) twice in succession, once for the $t$ and once
for the $\tbar$.  One new subtlety in this case is that the final
state contains two identical $\bbar$ antiquarks. One should therefore
antisymmetrize the total amplitude under the exchange of the two
$\bbar$'s. In practice, we implement cuts in such a way that the two
$\bbar$'s can effectively be distinguished.  In particular, in
$\tbbc$, we have $(p_b + p_\bbar + p_c)^2 = m_t^2$. But this relation
will not, in general, be satisfied if the $\bbar$ comes from the decay
of the $\tbar$. Thus, the two $\bbar$'s can be distinguished using
experimental cuts, and we therefore treat them as non-identical.
Further discussion on this point is included in the companion
paper~\cite{companion}.  Defining
\begin{eqnarray}
  \tilde{n}_{\tbar\mu} \equiv -\overline{\eta}_{\mu\nu}
  \mbox{Tr}\!\left[C\overline{C}
    \left(\pslash_{\tbar} - m_t\right)\gamma^5\gamma^\nu\right]/
     \mbox{Tr}\!\left[C\overline{C}\left(\pslash_{\tbar} 
       - m_t\right)\right],
\label{eq:ntbar}
\end{eqnarray}
where
\begin{eqnarray}
  \overline{\eta}_{\mu\nu} \equiv g_{\mu\nu} 
    - \frac{p_{\tbar\mu} p_{\tbar\nu}}{m_t^2},
\end{eqnarray}
and proceeding as above, we find~\cite{kawasaki}
\begin{eqnarray}
  && d\sigma\left(gg\to \left(b\bbar c\right)
    \left(\bbar\ell\nubar\right)\right) \nonumber\\
  && ~~~= \frac{4}{\Gamma_t^2}\sum_{b, \bbar, c \mbox{\scriptsize
~spins}}\,
     \sum_{\bbar, \ell, \nubar \mbox{\scriptsize ~spins}}
     d\sigma\left(gg\to t\left(n_t\right)
       \tbar\left(\tilde{n}_{\tbar}\right)\right)
     d\Gamma\left(t\to b \bbar c\right)
     d\Gamma\left(\tbar\to \bbar \ell\nubar\right) .~~~~~~
\label{eq:diffcrosssectiondoublespin2}
\end{eqnarray}
Use of the above expression requires some care, since the same subtle
issues are present as were noted above for the analogous expression in
Eq.~(\ref{eq:diffcrosssectionsinglespin}). 

\subsection{\boldmath Explicit Expressions for $n_t^\alpha$ and 
$\tilde{n}_{\tbar}^\alpha$}
\label{sec:diffwidth}
At this stage, let us work out expressions for the ``special'' $t$ and
$\tbar$ polarization four-vectors, $n_t^\alpha$ and $\tilde{n}_{\tbar}^\alpha$,
respectively. 
The quantity that is of interest in the calculation is
\begin{eqnarray}
 \sum_{b, \bbar, 
         c \mbox{\scriptsize ~spins}} 
    \!\!\!\mbox{Tr}\!\left[B \overline{B}\left(\pslash_t+m_t\right)\right] 
        n_t^\alpha 
  &=& -\eta^{\alpha\beta}\sum_{b, \bbar, 
         c \mbox{\scriptsize ~spins}} \mbox{Tr}\!\left[B{\overline{B}}
    \left(\pslash_t + m_t\right)\gamma^5\gamma_\beta\right] \nonumber\\
  & = &-\left(4\sqrt{2}G_F V_{tb}V_{cb}\right)^2
    \Bigg[\sum_{i,\sigma} 
      \frac{2\,p_i\cdot p_t}{m_t}\left(p_t^\alpha
    - \frac{m_t^2 p_i^\alpha}{p_i\cdot p_t}\right) \!\xi^\sigma \!A_i^\sigma \nonumber \\
   && ~~~~
    +32\,m_t
\,\mbox{Im}\left(X^T_{LL}X^{S*}_{LL}+X^T_{RR}X^{S*}_{RR}\right)
    \epsilon^{\alpha\beta\gamma\delta}p_{b\beta}
       p_{\bbar\gamma}p_{c\delta}\Bigg] .~~
\label{eq:dGam_ntalpha}
\end{eqnarray}
Note that this ``special'' polarization four-vector for the $t$ quark,
which will eventually be incorporated into the expression for
$\ggprocess$, contains all of the relevant information and
correlations related to the decay of the $t$.

Since the semileptonic decay of the $\tbar$ is assumed to be SM-like,
the expression for $\tilde{n}_{\tbar}^\alpha$ is much simpler.  Defining
\begin{eqnarray}
  A_\ell = \left(p_{\tbar} - p_{\ell}\right)^2 
    m_W^4 \left|G_T(2\,p_\ell\cdot p_{\overline{\nu}})\right|^2,
\end{eqnarray}
[in analogy with the SM part of Eq.~(\ref{eq:Abbar})], we find
\begin{eqnarray}
 \sum_{\bbar, \ell, \nubar \mbox{\scriptsize ~spins}} 
    \!\!\!\mbox{Tr}\!\left[C \overline{C}\left(\pslash_{\tbar}-m_t\right)\right] 
        \tilde{n}_\tbar^\alpha 
  &=& -\overline{\eta}^{\alpha\beta}\sum_{\bbar, \ell, \nubar \mbox{\scriptsize ~spins}} 
        \mbox{Tr}\!\left[C{\overline{C}}
    \left(\pslash_\tbar - m_t\right)\gamma^5\gamma_\beta\right] \nonumber\\
  & = &\left(4\sqrt{2}G_F V_{tb}\right)^2
      \frac{2\, p_{\ell}\cdot p_\tbar}{m_t}\left(p_{\tbar}^\alpha
    - \frac{m_t^2 p_\ell^\alpha}{p_{\ell}\cdot p_\tbar}\right)\!A_\ell \,.~~~
\label{eq:dGam_ntbaralpha}
\end{eqnarray}
Equations~(\ref{eq:dGam_ntalpha}) and (\ref{eq:dGam_ntbaralpha}) may
be compared to related expressions in Ref.~\cite{arens}.  The
following expressions are also useful:
\begin{eqnarray}
 \sum_{b, \bbar, 
         c \mbox{\scriptsize ~spins}} 
    \!\!\!\mbox{Tr}\!\left[B \overline{B}\left(\pslash_t+m_t\right)\right] 
  & = &\left(4\sqrt{2}G_F V_{tb}V_{cb}\right)^2
    \sum_{i,\sigma} 2\, 
      p_i\cdot p_t \,A_i^\sigma \,, \\
 \sum_{\bbar, \ell, \nubar \mbox{\scriptsize ~spins}} 
    \!\!\!\mbox{Tr}\!\left[C \overline{C}\left(\pslash_{\tbar}-m_t\right)\right] 
  & = &\left(4\sqrt{2}G_F V_{tb}\right)^2
      2\, p_{\ell}\cdot p_\tbar \,A_\ell \,.~~~
\end{eqnarray}
With these expressions in hand, we can now work out the final
expression for the differential cross-section.

\subsection{\boldmath Differential cross-section}
\label{sec:diffcsec}
Using Eqs.~(\ref{eq:dGam_ntalpha}) and (\ref{eq:dGam_ntbaralpha}) in
Eq.~(\ref{eq:diffcrosssectiondoublespin2}), we have
\begin{eqnarray}
  d\sigma\left(\ggprocess\right) =
    \left({\cal B}_{\mbox{\scriptsize non-TP}}
       + {\cal B}_{\mbox{\scriptsize TP}}\right)d\lambda ~,
\label{eq:dsigma2}
\end{eqnarray}
where
\begin{eqnarray}
\label{eq:Bnon-TP2}
{\cal B}_{\mbox{\scriptsize non-TP}} & = &
\sum_{i,\sigma} A_i^\sigma A_\ell
    \Big\{
    -\frac{p_i\!\cdot\! p_t \,p_\ell\!\cdot\! p_{\overline{t}}}{m_t^2}
    \left[f(r,z)\!+\!\xi^\sigma
\left(r^4\left(z^4-2\right)\!+\!1\right)\right] 
       \!-\! \xi^\sigma p_i\!\cdot\! p_\ell \, g(r,z) \nonumber\\
    && ~~~~~~~~~~~~~~-\frac{\left(r^2-1\right)
         \left[r^2\left(z^2-2\right)+1\right]
         \xi^\sigma}{2 m_t^2}
     \left(p_i\!\cdot\!Q \,Q\!\cdot\!p_\ell
         + p_i\!\cdot\!P_t \,P_t\!\cdot\!p_\ell\right)\nonumber\\
    && ~~~~~~~~~~~~~~-
    \frac{r^2(r^2-1)(z^2-1)\xi^\sigma}{2 m_t^2}\big[
      p_i\!\cdot\! P_g \left(
       P_g\!\cdot\! p_\ell - Q\!\cdot\! p_\ell\,rz
      \right) \nonumber\\
     && ~~~~~~~~~~~~~~~~~~~~~~~~~~~~~~~~~~~~~~~~~~+
      p_i\!\cdot\! Q \left(
       P_g\!\cdot\! p_\ell\,rz - Q\!\cdot\! p_\ell
      \right)
      \big]
    \Big\} ~, \\
{\cal B}_{\mbox{\scriptsize TP}} & = &
    16 A_\ell
\,\mbox{Im}\left(X^T_{LL}X^{S*}_{LL}+X^T_{RR}X^{S*}_{RR}\right)
    \Big\{ 
      -g(r,z)\epsilon\left(p_b, p_{\overline{b}},p_c,p_\ell\right)
      \nonumber\\
    && - \frac{\left(r^2-1\right)p_\ell\!\cdot\!p_{\overline{t}}}
      {m_t^2}\left[r^2\left(z^2-2\right)+1\right]
        \epsilon\left(p_b, p_{\overline{b}},p_c,Q\right)\nonumber\\
    && -\frac{r^2(r^2-1)(z^2-1)}{2 m_t^2}\big[
      \left(
       P_g\!\cdot\! p_\ell - Q\!\cdot\! p_\ell\,rz
      \right)\epsilon\left(p_b, p_{\overline{b}},p_c,P_g\right)
\nonumber\\
    && ~~~~~~~~~~~~~~~~~~~~~~~~+
      \left(
       P_g\!\cdot\! p_\ell\,rz - Q\!\cdot\! p_\ell
      \right)\epsilon\left(p_b, p_{\overline{b}},p_c,Q\right)
      \big]   
    \Big\} ~,
    \label{eq:BTP2}
\end{eqnarray}
and
\begin{eqnarray}
  d\lambda & = &
    \frac{\alpha_S^2 \,G_F^4 V_{tb}^4V_{cb}^2 \left(1-r^2\right) r }
      {4 \left(4\pi\right)^{10} \Gamma_t^2 \,m_t^2}
    \left(1-\frac{M_2^2}{m_t^2}\right)
    \left(1-\frac{M_5^2}{m_t^2}\right)
    \frac{(9r^2z^2+7)}{(r^2z^2-1)^2} \nonumber\\
    && \times dM_2^2 \,dM_5^2\,
d\Omega_1^{**}\,d\Omega_2^{*}\,d\Omega_4^{**}\,d\Omega_5^{*}\,d\Omega_t
~.
\label{eq:dlambda}
\end{eqnarray}
In the above, the $p_i$ are the momenta of the final-state quarks
coming from the top decay (i.e., $b$, $\bbar$ and $c$); also, $P_t$,
$Q$, $P_g$, $r$, $z$, $f(r,z)$ and $g(r,z)$ were defined in
Eqs.~(\ref{eq:Pt}), (\ref{eq:f}) and (\ref{eq:g}).  In arriving at the
above expression for $d\lambda$, we have decomposed the six-body phase
space into five solid angles and four invariant masses (see
Fig.~\ref{fig:kinematics}), and then have used the narrow-width
approximation for the $t$ and $\tbar$ quarks to eliminate two of the
invariant-mass degrees of freedom.  The solid angles
$d\Omega_1^{**}$-$d\Omega_t$ and the invariant masses $M_2$ and $M_5$
are discussed in Sec.~\ref{Sec:ggprocess}.

Inspection of Eqs.~(\ref{eq:Bnon-TP2}) and (\ref{eq:BTP2}) reveals
elements that are a combination of expressions coming from the
production and decay of the $t$ and $\tbar$ quarks.  The $A_i^\sigma$
are related to the decay $\tbbc$ [see
  Eqs.~(\ref{eq:Abbar})-(\ref{Ahatdefs})].  $A_\ell$ is similarly
related to the semileptonic decay of the $\tbar$.

\subsection{\boldmath Integrated cross-section}
\label{sec:intgcsec}
In the SM, 
\begin{eqnarray}
\sigma_{\mbox{\scriptsize SM}}
\quad &&\equiv \quad
\sigma\left(\ggprocess\right)\big|_{\mbox{\scriptsize SM}} \nonumber
\\[2ex]
\quad &&= \quad
\sigma \!\left(gg\to t\tbar\right) 
\mbox{BR}\!\left.\left(t\to b\bbar c \right)\right|_{\mbox{\scriptsize
SM}}
\mbox{BR}\!\left(\tbar\to \bbar\ell\nubar\right),
\label{eq:sigmaSM}
\end{eqnarray}
in which $\sigma \!\left(gg\to t\tbar\right)$ is defined in
Eq.~(\ref{eq:sigggttbar}), BR$\left.\left(t\to b\bbar c\right)
\right|_{\mbox{\scriptsize SM}}\! = \!V_{tb}^2V_{cb}^2/3$ and
BR$\left(\tbar\to \bbar\ell\nubar\right) \! =\!  V_{tb}^2/9$.

After the inclusion of new physics,
\begin{eqnarray}
\sigma_{\mbox{\scriptsize SM+NP}}
\quad &&\equiv \quad
  \sigma\left(\ggprocess\right) \big|_{\mbox{\scriptsize SM+NP}}
\nonumber\\[2ex]
  &&=\sigma_{\mbox{\scriptsize SM}}
    \Bigg\{
    1+ \frac{4\Gamma_W}{m_W}\mbox{Im}\left(X^{V*}_{LL}\right)
  + \frac{3 G_F m_t^2}
     {4\sqrt{2}\pi^2 \left(1-\zeta_W^2\right)^2
      \left(1+2\zeta_W^2\right)}
    \sum_{i,\sigma} \hat{A}_i^\sigma\Bigg\}.
\end{eqnarray}


\end{document}